\documentclass[a4paper,11pt]{article}
 \usepackage[margin=1in]{geometry}
\usepackage[square,numbers]{natbib}
\usepackage[utf8]{inputenc}
\usepackage{amsmath}
\usepackage{amsthm}
\usepackage{amsfonts}
\usepackage{bm}
\usepackage{thm-restate}
\usepackage{wrapfig}
\usepackage{subcaption}
\usepackage{float}
\usepackage{algorithm}
\usepackage[noend]{algorithmic}

\usepackage{color}   
\usepackage{hyperref}
\hypersetup{
    colorlinks=false, 
    linkcolor=blue,  
    urlcolor  = blue,
    linktoc=all,     
}

\usepackage[group-separator={,}]{siunitx}
\usepackage{framed}
\usepackage{setspace}

\usepackage{todonotes}

\theoremstyle{plain}

\theoremstyle{definition}

\theoremstyle{remark}

\usepackage{dsfont}

\newcommand{\remove}[1]{}

\newcommand{\theoSuggestion}[2]{#1}

\allowdisplaybreaks

\begin{document}
\pagenumbering{arabic}

\title{
Quantifying the sustainability impact of Google Maps: A case study of Salt Lake City\footnote{authors ordered alphabetically.}
}
\date{}

 \author{
 Neha Arora \\
 Google Research \\
 \small\texttt{nehaarora@google\!.\!com}
 \and
Theophile Cabannes\\
 Google Research\footnote{On leave from UC Berkeley.} \\
 \small\texttt{cabannes@google\!.\!com}
 \and
Sanjay Ganapathy\\
 Google Research \\
 \small\texttt{ganapathys@google\!.\!com}
 \and
Yechen Li\\
 Google Research \\
 \small\texttt{yechenl@google\!.\!com}
 \and
Preston McAfee\\
 Google Research \\
 \small\texttt{mcaf@google\!.\!com}
 \and
Marc Nunkesser\\
 Google Geo \\
 \small\texttt{marcnunkesser@google\!.\!com}
 \and
Carolina Osorio\\
 Google Research \\
 HEC Montreal \\
 \small\texttt{osorioc@google\!.\!com}
 \and
Andrew Tomkins\\
 Google Research \\
 \small\texttt{tomkins@google\!.\!com}
 \and
Iveel Tsogsuren\\
 Google Research \\
 \small\texttt{iveel@google\!.\!com}
 \and
}

\maketitle

\begin{abstract}
\theoSuggestion{
    Google Maps uses current and historical traffic trends to provide routes to drivers. In this paper, we use microscopic traffic simulation to quantify the improvements to both travel time and CO2 emissions from Google Maps real-time navigation. A case study in Salt Lake City shows that Google Maps users are, on average, saving 1.7\% of CO2 emissions and 6.5\% travel time. If we restrict to the users for which Google Maps finds a different route than their original route, the average savings are 3.4\% of CO2 emissions and 12.5\% of travel time. These results are based on traffic conditions observed during the Covid-19 pandemic. As congestion gradually builds back up to pre-pandemic levels, it is expected to lead to even greater savings in emissions.
}{
This article introduces a calibrated micro-simulation of Salt Lake City MONTH YEAR average daily traffic using SUMO and Google Maps internal data.
}

\end{abstract}

\section{Approach Overview}
We build a microscopic traffic simulation model of Salt Lake City, Utah, USA. Various well established traveler behavioral models for car-following behavior, and lane-changing behavior are used to model travel patterns. Metrics presented in this paper are computed from aggregated and anonymized data of Google users who have opted-in to Location History \cite{location_history}, a feature which is off by default. Travel demand is sourced from a combination of Location History data and city sensor vehicular count data \cite{udot_atspm}.
{To validate that trip durations in our simulation match the observed durations for those same trips in the real world, we compute aggregated and anonymized metrics based on Google Maps navigation data \cite{navlogs}.}
{The simulation is validated by comparing simulated trip durations with aggregated and anonymized trip durations from Google Maps navigation data \cite{navlogs}.
}
The traffic simulator is combined with an emissions model developed by the National Renewable Energy Lab (NREL) to yield emissions estimates \cite{nrel, nrel_journal}. 

To estimate travel time and emissions impacts of Google Maps real-time navigation, we begin with a baseline scenario in which vehicles follow the Location History based routes used in our high-fidelity simulation described above.  We then consider a routed scenario in which a fraction of vehicles, denoted the routed vehicles, are routed using real-time shortest paths as in Google Maps real-time navigation~\cite{map_guidance}. The number of routed vehicles is equal to the fraction of Google Maps users. We then evaluate the impact of real-time routing on emissions and trip duration for the routed vehicles by comparing the baseline and the routed scenarios. 

\subsection{Preserving Data Privacy}
As stated above, insights and metrics in this paper are derived from aggregated and anonymized sets of data from users who have turned on the Location History \cite{location_history} setting, which is off by default. People who have Location History turned on can choose to turn it off at any time from their Google Account \cite{google_account} and can always delete Location History data directly from their Timeline \cite{timeline}. Simulation validation metrics computed from Maps navigation data are also aggregated and anonymized. Access to this data is limited and audited. The non-anonymized data is encrypted at rest and models trained on this are not kept past the lifetime of the source data. 

To provide strong privacy guarantees, all reported metrics are anonymized and aggregated using a differentially private mechanism \cite{differential_privacy}. The automated Laplace mechanism adds random noise drawn from a zero mean Laplace distribution and yields $(\epsilon, \delta)$-differential privacy guarantee of $\epsilon = 1.09861$ and $\delta = 1e-5$ \cite{laplace_dist}.

To compute emission estimates the NREL model \cite{nrel, nrel_journal} was integrated into the simulator and the data never left Google servers.
\section{Simulation Framework}

The below figure summarizes the simulation framework. The main components are the traffic simulation combined with NREL's emissions model~\cite{nrel, nrel_journal}.

\subsection{Traffic Simulation} \label{sec:traffic-sim}
With the traffic simulation model we want to replay, via simulation, the traffic situation for a specified day as closely as possible. We use an open-source simulation software, SUMO (Simulation of Urban Mobility) \cite{sumo_2018}. We consider a single travel mode, which is a car, and a single vehicle type, which is SUMO’s default passenger car. To describe traffic dynamics, we use travel behavioral models available in SUMO. These describe, for instance, how drivers make decisions such as car-following, lane changing and speed limit compliance. Google Maps data is used to define the road network supply (e.g. network topology, connectivity) and the various attributes of each static road segment such as number of lanes, speed limit, and so forth. In this paper, we use the terms road and segment interchangeably.

\begin{figure}[h]
    \centering
    \includegraphics[width=\linewidth]{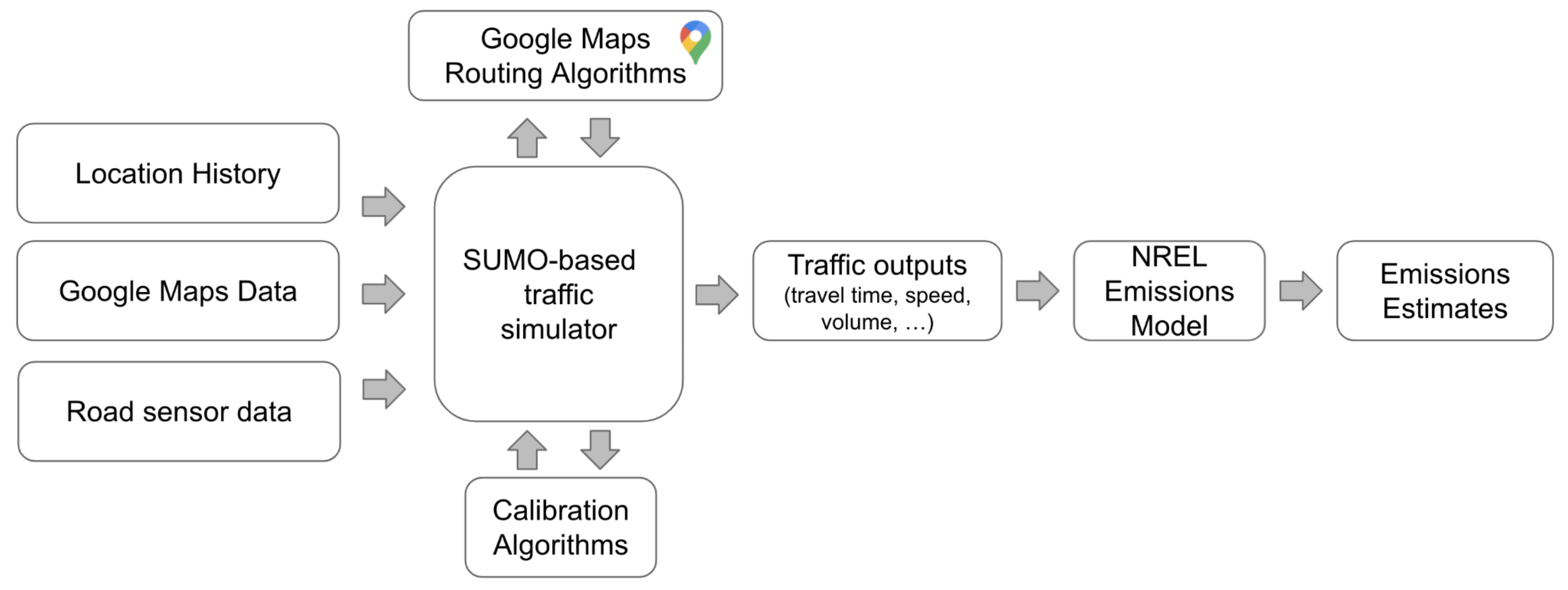}
    \caption{Simulation Framework.}
    \label{fig:iterationsComp}
\end{figure}

Calibration (i.e., input parameter estimation) and validation of the traffic simulator is carried out considering the baseline scenario. 
The vehicles (or trips) to be simulated are determined by a travel demand model. The road network of a metropolitan area is decomposed into zones defined by S2 cells of level 13~\cite{s2cells}, with an average area of 1.27 km$^2$. Travel demand for a given time period (e.g., one hour) is defined by a coarsened origin-destination matrix (or OD matrix) which gives the expected number of vehicles traveling from each origin zone to each destination zone. For a given OD pair, the set of feasible routes (i.e., the route choice set) in the baseline scenario is determined by the set of all routes derived from Location History data for the considered time period over several weeks.  

For a given time period, the entries in the OD matrix are obtained as a combination of Location History trajectories plus additional trajectories as required to match segment vehicular count data from city sensors \cite{udot_atspm}.
The additional trajectories are produced using the OD calibration algorithm described in~\cite{arora2021efficient}. This algorithm generates an OD matrix that induces simulated count data that is consistent with the overall city sensor segment count data.

For the Salt Lake City case study described here, the city sensor count data is obtained from the Utah Department of Transportation (UDOT) Automated Traffic Signal Performance Metrics (ATSPM) portal \cite{udot_atspm}. We measure the error between ground truth data from the portal versus a particular OD matrix through the count-normalized root mean square error (nRMSE):
\begin{equation}
    nRMSE = 100*\frac{\sqrt{\frac{1}{|I|} \sum_{i \in I} {(y_i - \hat{y}_i)^2}}}{\frac{1}{|I|} \sum_{i \in I} {y_i}}
\label{eq:nRMSEDef}
\end{equation}
where $I$ denotes the set of segments where we have ground truth, $y_i$ is the ground truth count of segment $i$, and  $\hat{y}_i$ denotes the calibration-based estimate of $y_i$. 
For this case study, calibrated OD matrices are obtained by considering 415 segments with ground truth data. They yield  nRMSE's of $22 \pm 2 \%$ for off-peak hours and $35 \pm 2 \%$ for peak hours.

To validate the quality of the calibrated ODs, and more generally the  ability of the traffic simulator to accurately represent real traffic patterns, we compare the simulated trip travel times (i.e., simulated ETAs) to those observed from Google Maps navigation logs. For the current case study, the nRMSE of the simulated ETAs is $25 \pm 1\%$ for peak hours and $17 \pm 1\%$ for off-peak hours. The definition of nRMSE is similar to Equation~\eqref{eq:nRMSEDef}, where we replace the set of ground truth counts with ground truth trajectory durations.

\begin{table}
\begin{center}
    \begin{tabular}{|c|c|c|}
    \hline
    Hour of Day  &  ETA nRMSE (\%) & Segment Counts nRMSE (\%) \\
    \hline
    11 am (Off peak) & $17 \pm 1\%$ & $22 \pm 2 \%$   \\
    5 pm (Peak) & $25 \pm 1\%$ &$35 \pm 2 \%$    \\
    \hline
    \end{tabular}
\end{center}
\caption{Summary of simulation evaluation in Salt Lake City}
\label{tab:nRMSE}
\end{table}

\subsection{Emissions Model}

We use NREL’s Route Energy Prediction model, RouteE \cite{nrel, nrel_journal}, that predicts the energy consumption and emissions for any given route. This model accounts for driving conditions such as traffic speed, road grade and turns. For every ten minutes of simulation, we produce a snapshot of the current traffic on each road segment. Based on this snapshot, we then estimate the fuel consumption and emissions for each simulated trip using the RouteE model.

\subsection{Experimental setup}
To evaluate the impact of Google Maps route guidance on CO2 emissions, we compare the baseline scenario with the routed scenario. The baseline scenario is intended to capture vehicles who are making their own route choices, as described above in Section~\ref{sec:traffic-sim}.

The routed scenario considers the same set of vehicles, but replaces the route choice of a subset of those with the shortest path based on real-time traffic information, as in Google Maps with real-time navigation. The vehicles re-routed in this way (called the \emph{routed vehicles}) are chosen uniformly at random from all the vehicles. The rest of the vehicles are referred to as \emph{background vehicles}. The background vehicles and their corresponding routes remain identical to those of the baseline scenario. The number of routed vehicles corresponds to the number of navigation requests received for Google Maps travel guidance during the same time period.

As described in Section~\ref{sec:traffic-sim},  the route choice set of the baseline scenario consists of routes observed from Location History data. If a baseline vehicle is already using a navigation service for real-time routing, that vehicle may already be on the best route, and our simulation analysis may understate the improved outcomes that are possible using real-time traffic information.

For the routed scenario, the real-time routes provided to routed vehicles are selected as follows.
For each origin-destination pair, we query Google Maps to obtain a set of plausible routes for the given source/destination pair. Next, we maintain a snapshot of current simulated travel time on each road segment, updated every ten minutes, and use the most recent snapshot to score and rank the routes based on travel time.  Finally, we select the fastest route, and do not allow any changes to the route during driving. In other words, only pre-trip real-time guidance is considered, en-route guidance is not accounted for. Again, this may underestimate the congestion impacts of the case study.  
To estimate the impact of Google Maps real-time navigation we compare the travel time and CO2 emissions for each routed vehicle in the baseline versus the routed scenario.
\section{Case Study Results}
We build the network for a region of Salt Lake City with 45K road segments.
We consider a representative day in December 2020, during the Covid-19 pandemic and analyze hourly simulation runs for the hours from 7am to 7pm. For each hour of the day we simulate between 82,500 to 137,000 trajectories per hour across 12,434 OD pairs distributed throughout the network. The average driving distance is approximately 6km.

For each hour, we run 6 simulation replications. Sources of stochasticity across the replications include sampling from the demand distribution, and sampling the set of routed trips. In the analysis of travel time and CO2, we exclude unrealistic trips that have a duration that is more than twice that of the shortest path or is longer than 20 minutes. We do this in order to limit the potential to overestimate the travel time and CO2 impacts.

\begin{figure}[h]
    \centering
    \includegraphics[width=\linewidth]{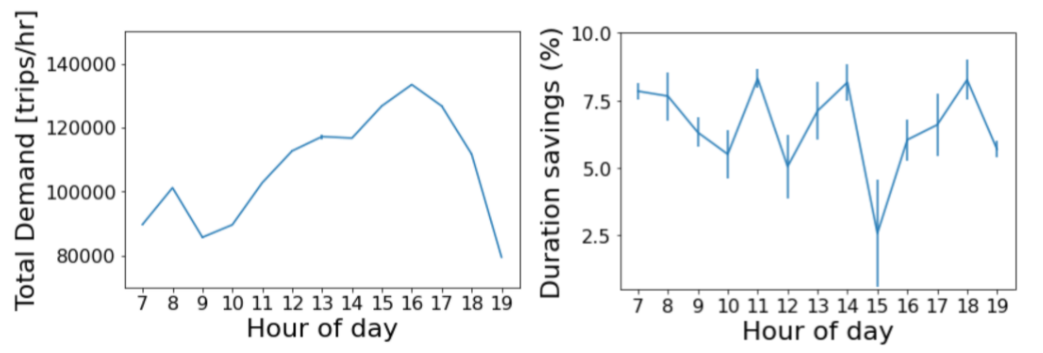}
    \caption{Total hourly demand (left plot) and percentage savings in duration (right plot). Error bars show standard deviation across replications.}
    \label{fig:demDuration}
\end{figure}

The plot on the left of Figure~\ref{fig:demDuration} shows hourly total demand from 7am to 7pm in Salt Lake City for that day. There are error bars that have a half-width equal to the standard deviation. However, the error bars are barely visible in this plot.
Demand gradually increases throughout the day, with a peak period between 4pm and 6pm. Interestingly, we observe no morning peak period but only afternoon peak period. This is consistent with recent analysis of other cities carried out during the pandemic ~\cite{streetlight_covid}.

The plot on the right of Figure~\ref{fig:demDuration} shows the average, across all routed vehicles, of the percentage difference between the baseline-scenario trip duration and the routed-scenario trip duration. A positive percentage change indicates a reduction in trip duration after routing using real-time traffic information. The error bars are computed in the same way as for the left plot. The average saving varies between $2.5\%$ and $7.7\%$ over the day.

\begin{figure}[h]
    \centering
    \includegraphics[width=\linewidth]{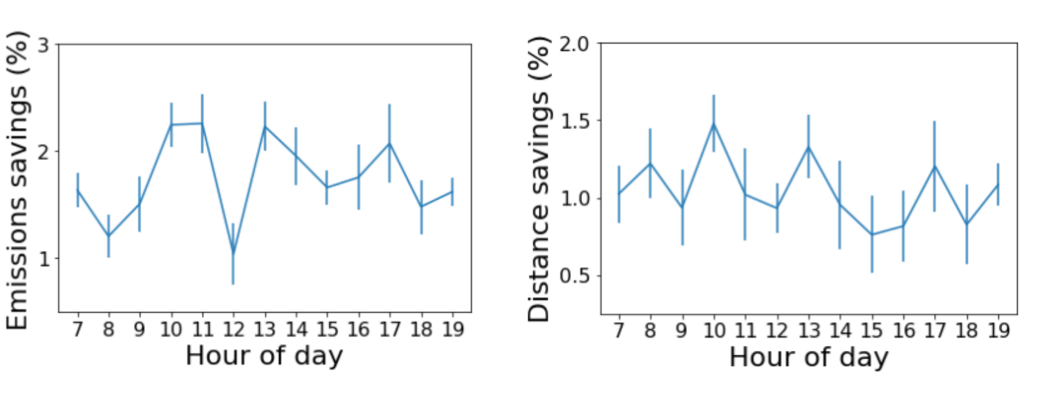}
    \caption{Percentage savings in emissions (left plot) and distance (right plot). Error bars show standard deviation across replications.}
    \label{fig:emissionsDistance}
\end{figure}

Figure~\ref{fig:emissionsDistance} shows trip emissions (left) and trip distance (right). As for duration, the $y$ axis shows the average across all routed vehicles of the percentage difference between the baseline-scenario and the routed-scenario trip.
The left plot indicates that the average percentage emissions savings vary, throughout the day, between $1\%$ and $2.2\%$. Similarly, the right plot indicates that the average percentage distance savings vary, throughout the day, between $0.8\%$ to $1.4\%$.
For most of the day, emissions savings are positively correlated with distance savings. Routed vehicles save $6.5\%$ travel duration on average and $2.6\%$ to $8.3\%$ depending on the time of day. Conditioning on routed vehicles for whom the route actually changed, so they were not already on the shortest path, these vehicles save $3.4\%$ of emissions and $12.5\%$ of travel time on average.

Our analysis shows that the emission savings are mostly coming from routing users more efficiently to an optimal route with less travel duration and travel distance, both for long trips and short trips. While in principle the shortest duration path could result in higher travel distance and emissions, our analysis shows that in practice, careful routing is a win-win-win, saving simultaneously on time, distance, and emissions.

\section{Discussion}
This paper develops a high-resolution disaggregate traffic and emissions simulation framework for metropolitan vehicular traffic. The framework is used to quantify the travel time and CO2 emissions impacts of real-time navigation with Google Maps. Location History, Google Maps and UDOT data are combined to calibrate and validate the vehicular traffic model. The simulation is combined with NREL’s RouteE model to estimate vehicular CO2 emissions. The analysis focuses on Salt Lake City. It compares two scenarios that differ in whether or not a subset of vehicles are routed with Google Maps. Routed drivers save $6.5\%$ travel time and $1.7\%$ of emission by Google Maps navigation.
The results are based on traffic conditions observed during the Covid-19 pandemic, where lower levels of congestion are observed. The impact of real-time navigation is expected to be  greater for higher levels of congestion. The routing mechanism of this paper does not account for en-route trip guidance, which is commonly used in Google Maps. We expect en-route guidance to be particularly important and impactful for congested scenarios. 
This work is part of ongoing work that quantifies the sustainability impact of real-time route guidance. This paper focuses on the common policy of routing based on shortest travel time. Ongoing work considers guidance based on both travel time and emissions costs. It contributes to advancing our understanding of the tradeoffs between travel time and emissions. Moreover, it informs the design of future real-time navigation services that improve both network efficiency and sustainability.
\bibliography{generated_main}

\end{document}